\documentclass[11pt]{article}

\usepackage[final]{acl}
\usepackage{hyperref}
\usepackage{url}
\usepackage{times}
\usepackage{latexsym}
\usepackage{url}            %
\usepackage{wrapfig}
\usepackage{enumitem}
\usepackage{booktabs}       %

\usepackage{caption}
\usepackage{placeins}
\usepackage{amssymb}%
\usepackage{pifont}%
\usepackage{floatrow}
\usepackage{fancyvrb}
\usepackage{graphicx}
\usepackage{amsmath}
\usepackage{amsfonts}       %
\usepackage{nicefrac}       %
\usepackage{microtype}      %
\usepackage{xspace}         %
\usepackage{multirow}
\usepackage{array}
\usepackage{siunitx}
\usepackage{booktabs}
\usepackage{colortbl}
\usepackage{color}
\usepackage{soul}
\usepackage[many]{tcolorbox}
\usepackage{listings}
\usepackage{hyperref}
\usepackage{fancyvrb}

 \usepackage{textcomp}
 \newcommand{\mytexttilde}{\raisebox{0.5ex}{\texttildelow}}
 
\def\Snospace~{\S{}}

\newcommand\extrafootertext[1]{%
    \bgroup
    \renewcommand\thefootnote{\fnsymbol{footnote}}%
    \renewcommand\thempfootnote{\fnsymbol{mpfootnote}}%
    \footnotetext[0]{#1}%
    \egroup
}

\usepackage[T1]{fontenc}
\definecolor{mygray}{RGB}{220, 220, 220}
\definecolor{lightgreen}{RGB}{223,255,219}
\definecolor{lightred}{RGB}{255,219,219}

\definecolor{darkpink}{rgb}{0.91, 0.33, 0.5}

\definecolor{darkmagenta}{rgb}{0.55, 0.0, 0.55}
\definecolor{pos}{RGB}{167, 199, 231}
\definecolor{neg}{RGB}{250, 160, 160}
\definecolor{brickred}{rgb}{0.8, 0.25, 0.33}
\usepackage{soul}

\usepackage[utf8]{inputenc}

\newcommand{\modelname}{Promptriever\xspace}

\title{Promptriever: Instruction-Trained Retrievers \\ Can Be Prompted Like Language Models}

\author{
    \textbf{Orion Weller}$^{\hspace{.1em}
    {\color{brickred}\boldsymbol{\ast}}
    \hspace{.1em}{\color{blue}\boldsymbol{\iota}}}$    
    \quad
    \textbf{Benjamin Van Durme}$^{\hspace{.1em}\color{blue}\boldsymbol{\iota}}$
    \quad
    \textbf{Dawn Lawrie}$^{\hspace{.1em}\color{blue}\boldsymbol{\iota}}$ \\
    \quad
    \textbf{Ashwin Paranjape}$^{\hspace{.1em}\color{blue}\boldsymbol{\alpha}}$
    \quad
    \textbf{Yuhao Zhang}$^{\hspace{.1em}\color{blue}\boldsymbol{\alpha}}$
    \quad
    \textbf{Jack Hessel}$^{\hspace{.1em}\color{blue}\boldsymbol{\alpha}}$
    \vspace{.5em}\\
    $^{\color{blue}\iota\hspace{.1em}}$Johns Hopkins University
    \quad
    $^{\color{blue}\alpha\hspace{.1em}}$Samaya AI
     \vspace{.5em}\\
    \texttt{oweller@cs.jhu.edu}
}

\begin{document}
\maketitle
\begin{abstract}
Instruction-tuned language models (LM) are able to respond to imperative commands, providing a more natural user interface compared to their base counterparts.
In this work, we present Promptriever, the first \emph{retrieval} model able to be prompted like an LM.
To train \modelname,
we curate and release a new instance-level instruction training set from MS MARCO \cite{msmarco}, spanning nearly 500k instances. \modelname not only achieves strong performance on standard retrieval tasks, but also follows instructions. We observe:
(1) large gains (reaching SoTA) on following detailed relevance instructions (+14.3 p-MRR / +3.1 nDCG on FollowIR), (2) significantly increased robustness to lexical choices/phrasing in the query+instruction %
(+12.9 Robustness@10 on InstructIR), and
(3) the ability to perform hyperparameter search via prompting to reliably improve retrieval performance %
(+1.4 average increase on BEIR). 
\modelname demonstrates that retrieval models can be controlled with prompts on a per-query basis, setting the stage for future work aligning LLM prompting techniques with information retrieval.\extrafootertext{{\color{brickred}$^\ast$} Work performed during an internship at Samaya AI.}\footnote{Code and data are available at \url{https://github.com/orionw/promptriever}}

\end{abstract}

\section{Introduction}

\begin{figure}[t]
    \centering
    \includegraphics[width=0.9999\columnwidth,trim=0.5cm 0.5cm 1.0cm 0cm]{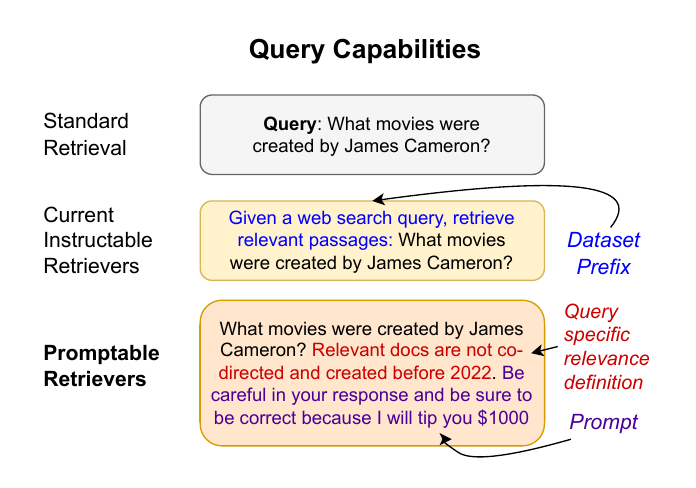}

    \caption{
    An illustration of the capabilities of retrieval models. Standard retrieval models find semantic similarity to the input query, typically matching using query keywords and phrases. Current instructable retrievers prepend a \textcolor{blue}{dataset prefix} that generically describes the task and is also used in training. We propose \textit{promptable retrievers} which can handle complex instructions including \textcolor{red}{detailed relevance definitions} and \textcolor{darkmagenta}{zero-shot prompting techniques} that act as a form of zero-shot hyperparameter optimization, similar to prompting LMs. 
    \vspace{-1em}
    }
    \label{fig:teaser}
\end{figure}

Modern information retrieval (IR) models generally
match queries to passages based on a single semantic similarity score.
As a result, the user experience of search can be opaque, with users needing to find particular keywords/phrasings, apply various filters in advanced search settings, and iterate based on previous searches
to find the ``just right" query that returns the desired passages.

In this work, we introduce \textit{\modelname}: a retrieval model that can instead be controlled via natural language prompts.
For example, if a user is searching for James Cameron movies, but is only interested in movies prior to 2022 that are not co-directed,
instead of applying a series of searches/hard filters, \modelname can adjust its notion of relevance dynamically based on a natural language description:
``Relevant documents are not co-directed, and are created before 2022" (Figure~\ref{fig:teaser}).

\begin{figure*}[t]
    \centering    \includegraphics[width=0.9999\columnwidth,trim=0.5cm 0cm 0.0cm 0.5cm]{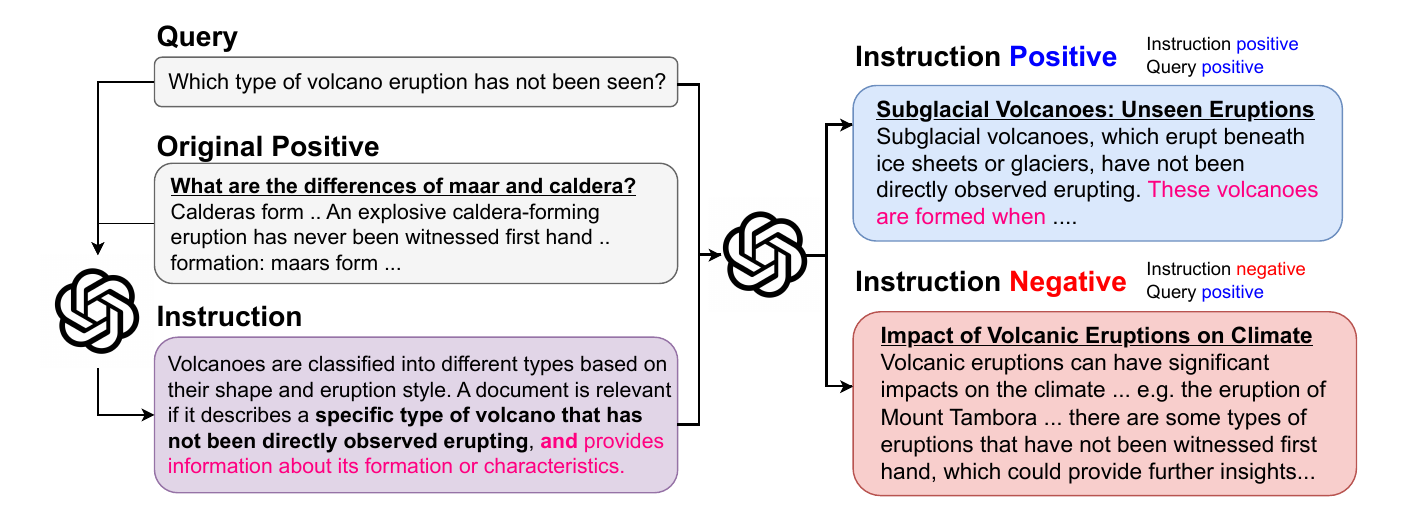}

    \caption{
    The data generation process to generate instruction-based retrieval data. We take the initial query and relevant passage and prompt an LM to generate an instruction that would match that query. Note that the instruction adds \textcolor{darkpink}{extra qualifications to the definition of relevance}. We then ask an LM to generate an example relevant and non-relevant passage for that query \textit{and} instruction. We see that the generated positive passage fulfills the extra requirement (in \textcolor{darkpink}{pink}) but the generated instruction-negative does not. We generate multiple types of instructions (both in length and style) for training set diversity.
    \vspace{-1em}
    }
    \label{fig:example}
\end{figure*}

\modelname is a bi-encoder retriever; its backbone is a large language model (LM) such as LLaMA-2 7B \cite{touvron2023llama,touvron2023llama2}.\footnote{We experiment with multiple backbones in Section~\ref{sec:backbones}.} %
Before IR training, language %
models
can readily adapt their outputs based on natural language (also referred to as \textit{instructions} or \textit{prompts}).
But after standard IR training, which typically focuses on
optimizing a single query-passage ``semantic similarity" score,
 instruction following capacity is not maintained (see  \S\ref{sec:ablations}). While some recent work 
has added instructions to retrieval model training sets \cite{instructor_models,asai2022tart,Jiang2023Mistral7,muennighoff2024generative} these are templated ``instructions," prepended to every query in the dataset at training and test time, rather than language defining relevance conditions on a per-instance basis. For example, when using the MS MARCO dataset, %
the standard ``instruction" is \textit{``Given a web search query, retrieve relevant passages that answer the query"} which is prepended to every query
\cite{muennighoff2024generative,wang2023improving,moreira2024nvretrieverimprovingtextembedding}.

\modelname, in contrast, is trained to maintain the per-instance instruction following capability of its backbone LM. To do so, we curate and release a synthetic dataset of \mytexttilde500K query-passage relevance pairs augmented with instance-level instructions.
The two key novelties in the training set are: (1) instructions defining per-query relevance in diverse, free-form natural language; and (2) %
instance-level \textit{``instruction negatives."} %
Instruction negatives are cases where the (query, passage) pair is highly relevant if viewed in isolation, but, the addition of a carefully constructed instruction significantly decreases that relevance. For example, a generated instruction can request additional, fine-grained information not present in a topically relevant passage, as in Figure~\ref{fig:example}. By construction, to achieve low training loss on instruction negatives, models must adapt their notion of relevance per-query based on the instruction.

\modelname not only achieves strong retrieval performance in standard settings, but also follows instructions more effectively than prior models. \modelname achieves:
\begin{itemize}[leftmargin=*,topsep=0pt,itemsep=-1ex,partopsep=1ex,parsep=1ex]
    \item State-of-the-art bi-encoder performance on instruction-following retrieval tasks (+14.3 p-MRR, +3.1 nDCG/MAP on FollowIR \citep{weller2024followir}) with comparable performance to SoTA cross-encoders.
    \item Improved robustness to query length/phrasing compared to RepLLaMA \cite{ma2023fine} with a 44\% decrease in variance on BEIR \cite{thakur2021beir} across instructions and a +12.9 improvement on InstructIR's \cite{oh2024instructir} Robustness@10 metric.
    \item Reliable improvements in retrieval performance zero-shot solely by prompting (such as adding \textit{``Think carefully when assigning relevance and I will give you a tip"}), enabling prompt engineering efforts and auto-prompting methods.
\end{itemize}

Our results demonstrate that with the right training data, modern bi-encoders \textit{can} be instructed/prompted in free-form natural language, in a similar manner to LMs. We hope this alignment between the LM and IR communities will allow for further improvements to dense retrieval models.

\section{\modelname Data Generation}
To train a bi-encoder that can retrieve based on instructions, %
we start with the popular IR training dataset, MS MARCO \cite{msmarco}. We use the \texttt{tevatron-msmarco-aug} version which includes hard-negatives and was used to train RepLLaMA \cite{ma2023fine}.
This training set includes roughly 491k queries and provides a positive passage and 30 hard negatives for each. We augment this set with instructions in a two part process: (1) instruction generation from the initial queries and (2) instruction-negative mining. These processes are illustrated in Figure~\ref{fig:example}. Our full dataset can be found on Huggingface at \url{https://huggingface.co/datasets/samaya-ai/msmarco-w-instructions}.

\subsection{Instruction Generation}
We start by generating an instruction for each query in MS MARCO. Consider the example in Figure~\ref{fig:example} where the initial query (``\textit{Which type of volcano eruption as not been seen?}") is seeking just a \textit{type of volcano.}
We ask an LM to generate instructions that: %
add additional requirements, explicitly exclude certain types of passages, or use ambiguity to make the initial query more specific.
For example, in the volcano query, 
the generated instruction asks for both the volcano type \textit{and} information about its formation (in \textcolor{darkpink}{pink}).
We use Llama-3-70B-Instruct \cite{llama3modelcard} to generate these more specific instructions at scale.\footnote{See Appendix~\ref{app:prompts} for prompt details}

\paragraph{Instruction diversity}
To ensure a wide range of instructions, we ask Llama 3 to generate instructions of:
1) varying length formats (from short one-sentence instructions to two paragraph-sized\footnote{While we don't expect real users of IR systems to regularly type two-paragraph instructions, including these helps the model learn diversity of length and adapt to (potentially machine generated) diverse or specific requests.} instructions), and 2) differing ``styles", which can either be a persona of the person giving the query, negation of some aspect, or generic background information. The generated volcano instruction in Figure~\ref{fig:example} has the generic background ``style" and was requested to be two sentences long.

Overall, Llama 3 succeeded in following length+style specifications (Table~\ref{tab:statistics}). A qualitative exploration of examples is in the Appendix: by style in Table~\ref{tab:instruction_examples}, and by length in Table~\ref{tab:length_examples}. %

\paragraph{Maintaining original MS MARCO positive relevance}
We aim to generate instructions that maintain the positive relevance relation between (query, passage) pairs from MS MARCO.
We provide both the query \emph{and positive passage} to the LM when generating instructions, and request that the more specific instruction keep the passage relevant. To check for success in this regard: we use FollowIR-7B \cite{weller2024followir}, a cross-encoder capable of making nuanced relevance judgments regarding (query, instruction, passage) instances. FollowIR-7B marked roughly 15\% of the generated instructions as making the original positive passage no longer relevant. In these cases, we substitute the original positive document with one generated from the next stage.\footnote{We evaluated 20 of these instructions that passed the filter and found all positive passages were still relevant.}

\subsection{Instruction Negative Mining}
After generating the instructions, it's possible to train models using the exact same data as RepLLaMA, except the queries have been augmented with instructions. However, our instructed-augmented queries have not changed any relevance relations in the corpus: with no additional modification, models could simply \emph{ignore} our instructions entirely, and achieve the same performance.\footnote{As we show as an ablation in Table~\ref{tab:instruct_ablate}, in this setting, models indeed do ignore the instruction.}  %

\begin{table}[t!]
\centering
\begin{tabular}{ll|ccc}
\toprule
& \textbf{Category} & \textbf{Min} & \textbf{Mean} & \textbf{Max} \\
\midrule
\parbox[t]{3mm}{\multirow{3}{*}{\rotatebox[origin=c]{90}{
\parbox[c]{2cm}{\centering \scriptsize Style Feature}}}} & None & 16 & 101 & 369 \\
& Negation & 18 & 98 & 363  \\
& Background & 19 & 108 & 340  \\
& Persona & 26 & 106 & 374 \\
\midrule
\parbox[t]{3mm}{\multirow{3}{*}{\rotatebox[origin=c]{90}{
\parbox[c]{2cm}{\centering \scriptsize Length Format }}}} & Short & 16 & 40 & 84  \\
& Medium & 43 & 84 & 154  \\
& Long & 43 & 107 & 185 \\
& Very Long & 96 & 181 & 374  \\
\midrule
& All & 16 & 103 & 374 \\
\bottomrule
\end{tabular}
\caption{Word-level dataset statistics from augmenting MS MARCO train with instructions. The dataset was generated from a cross-product of length format and features with each subset having approximately 122k instances (and the total dataset \mytexttilde490k). The mean numbers are rounded to the nearest digit. We see that LLama 3 70B generally followed the length description.\vspace{-1em}}
\label{tab:statistics}
\end{table}

Thus, we develop a complementary data augmentation that encourages models to pay attention to the instruction. We term this augmentation \textit{instruction negatives:}\footnote{This is similar to the intuition of \citet{asai2022tart} but crucially the instruction-negatives are gathered at an instance level rather than at the dataset-level.}  where the passage is query-positive but instruction-negative, i.e., when the instruction is added it decreases the passage's relevance. To achieve low training loss, the model must learn to condition on both query \textit{and} instruction.

Initial attempts to gather instruction negatives from the MS MARCO collection itself fell short: qualitatively, we found that the corpus does not contain enough positive passages per query to mine nuanced query-positives but instruction-negatives. Thus, we turn to generating passages with a LM.

We use \texttt{gpt-4o-2024-05-13} to generate the instruction negative passages, generating one query-positive/instruction-positive passage and three query-positive/instruction-negative passages per (query, instruction) pair.
We over-generate candidates and then filter them post-hoc %
because %
initial testing revealed that (on average) only two out of three generated passages were correctly query-positive/instruction-negative.\footnote{We found that current LMs struggle with this nuance and that only the most capable LMs (e.g. GPT-4o but not GPT-4o-mini) were able to produce instruction negatives with high enough efficiency to be practical. As described in the previous section, we use the query-positive/instruction-positive passage as the backup positive passage if the original positive passage was not relevant to the generated instruction.} We again use FollowIR-7B for the filter by: 1) checking that the generated instruction negatives are actually instruction-negative (and discarding it if not) and 2) checking that the generated instruction-positive was actually relevant (and discarding if not).

\paragraph{Filtering validation} %
We tasked 4 human annotators with the filtration task: %
for a given (query, instruction, generated passage) triplet, is the passage \textit{relevant} or \textit{not} to the query+instruction.
For this task, average human-human agreement was 75\% (N=32), whereas, the average human-model agreement was 84\% (N=64).
Thus, %
FollowIR-7B acts as a sufficiently high-quality filter. %

\newcommand{\graytext}[1]{\textcolor{gray}{#1}}
\begin{table*}[t!]
\centering
\resizebox{\textwidth}{!}{%
\begin{tabular}{ll|cc|cc|cc|cc|cc}
\toprule
 & \multirow{3}{*}{Model} & \multicolumn{8}{c|}{FollowIR} & \multicolumn{2}{c}{InstructIR}  \\
 &  & \multicolumn{2}{c|}{Robust04} & \multicolumn{2}{c|}{News21} & \multicolumn{2}{c|}{Core17} & \multicolumn{2}{c|}{Average} & \multicolumn{2}{c}{MS MARCO}  \\
&  & MAP & p-MRR & nDCG & p-MRR & MAP & p-MRR & Score & p-MRR & nDCG & Robust. \\
 \midrule
\parbox[t]{5mm}{\multirow{3}{*}{\rotatebox[origin=c]{90}{
\parbox[c]{1cm}{\centering \scriptsize Cross \\ Encoders}}}} & MonoT5-3B & \textbf{27.3} & +4.0 & 16.5 & +1.8 & 18.2 & +1.8 & 20.7 & +2.5 & - & - \\
&  Mistral-7B-instruct & 23.2 & +12.6 & 27.2 & +4.8 & 19.7 & +13.0 & 23.4 & +10.1 & 63.1 & 35.3 \\
& FollowIR-7B & 24.8 & \textbf{+13.7} & \textbf{29.6} & \textbf{+6.3} & \textbf{20.0} & \textbf{+16.5} & \textbf{24.8} & \textbf{+12.2} & \textbf{81.3} & \textbf{71.5} \\

\midrule
\parbox[t]{5mm}{\multirow{7}{*}{\rotatebox[origin=c]{90}{
\parbox[c]{2cm}{\centering \scriptsize Bi Encoders}}}} & BM25 & 12.1 & -3.1 & 19.3 & -2.1 & 8.1 & -1.1 & 13.2 & -2.1 & 76.0 & 26.9 \\
& TART-Contriever & 14.3 & -9.0 & 21.8 & -3.0 & 13.3 & -3.0 & 16.5 & -5.0 & 84.8 & 47.5 \\
& Instructor XL & 19.7 & -8.1 & 26.1 & -0.9 & 16.8 & 0.7 & 20.9 & -2.8 & 48.6 & 21.5 \\
& E5-Mistral & 23.1 & -9.6 & 27.8 & -0.9 & 18.3 & +0.1 & 23.1 & -3.5 & 86.3 & 55.4 \\
& Google Gecko & 23.3 & -2.4 & \textbf{29.5} & +3.9 & \textbf{23.2} & +5.4 & 25.3 & +2.3 & - & - \\
 & RepLLaMA & 24.0 & -8.9 & 24.5 & -1.8  & 20.6 & +1.3 & 23.0 & -3.1 & 85.7 & 50.2 \\
& \modelname & \textbf{28.3} & \textbf{+11.7} & 28.5 & \textbf{+6.4} & 21.6 & \textbf{+15.4} & \textbf{26.1} & \textbf{+11.2} & \textbf{92.1} & \textbf{63.1} \\

\bottomrule
\end{tabular}
}
\caption{Results for instruction following on the FollowIR and InstructIR datasets.  Higher is better for all metrics; MAP@1000/NDCG@5/Robustness@10 range from 0-100; p-MRR ranges from -100 to 100. Despite using the same backbone model (Llama  2) \modelname significantly outperforms RepLLaMA with a +3.1 gain on the standard retrieval score (nDCG/MAP average) and a +14.3 point gain on p-MRR. \textbf{\modelname outperforms all other dense retriever models and scores comparably to the best cross-encoder}, FollowIR-7B, despite not using attention between query and documents. Gecko scores use a proprietary API and were not reported for InstructIR. Bold results indicate the best for that architecture type (e.g. cross-encoder, bi-encoder).\vspace{-1em}}
\label{tab:instruct}
\end{table*}

\section{Experimental Settings}
Our goal is to show the eficacy of instruction-training in IR compared to standard training. Thus, we primarily compare to RepLLaMA and use their data/recipe for apples-to-apples comparison. 

\subsection{Training}
We train \modelname on the RepLLaMA MS MARCO data as well as the new instruction data generated by Llama 3 and GPT-4o. We use the same learning rate and other hyperparameter details as the original RepLLaMA for a fair comparison (see Appendix~\ref{app:hyperparameters} for more details).\footnote{Note that although this means that \modelname has seen more data overall, we provide comparisons in Section~\ref{sec:ablations} controlling for training data volume. Our findings do not change (i.e., that \modelname's instruction following capacity is not simply a result of seeing more datapoints).} We use all valid instruction-negatives in training and sample the remainder of the hard-negatives from the dataset used to train RepLLaMA (keeping the same number of hard negatives per query).

\subsection{Evaluation Datasets}

We evaluate on in-domain (MS MARCO), out-of-domain \citep[BEIR]{thakur2021beir}, and instruction-following retrieval datasets, including InstructIR \citep{oh2024instructir} and FollowIR \citep{weller2024followir}. Note that these instruction-following datasets evaluate instructions on a \textit{per-query} basis.

The metrics used are nDCG@10 for BEIR, TREC DL19 \cite{craswell2020overview} and DL20 \cite{soboroff2021overview}, and MRR for MS MARCO Dev.

The instruction-following datasets use both standard and instruction-following metrics: nDCG@5 for the News21 portion of FollowIR, MAP@1000 for the Core17 and Robust04 portions, as well as using p-MRR for all portions (ranging from -100 to 100) which measures the sensitivity to instructions in the prompt (-100 means it follows the opposite of the instruction, 0 means no change, and 100 means perfect instruction following).  InstructIR uses nDCG@10 as well as Robustness@10 which measures the minimum nDCG@10 score over 10 different prompts. Higher is better for all metrics.

We primarily compare with RepLLaMA in order to have an apples-to-apples comparison. However, we also show results for a variety of other models, including MonoT5 \cite{nogueira2020document}, Instructor models \cite{instructor_models}, the bi-encoder TART model trained from Contriever \cite{asai2022tart}, E5 Mistral \cite{wang2023improving}, Google Gecko \cite{lee2024gecko}, and BM25 \cite{robertson1995okapi}.

\section{Results}
\modelname outperforms the original RepLLaMA in instruction following (\S\ref{sec:sec_instruction_following}) while maintaining strong standard retrieval performance (\S\ref{sec:sec_standard_retrieval}). We also demonstrate that \modelname can be reliably zero-shot prompted, in the same manner as a language model (\S\ref{sec:sec_with_few_shot_prompting})

\subsection{Instruction Following}
\label{sec:sec_instruction_following}
Table~\ref{tab:instruct} presents the results for the FollowIR and InstructIR datasets. \modelname is the highest performing dense retriever, improving over RepLLaMA by +14.3 p-MRR (-3.1 $\to$ +11.2) and +3.1 in nDCG/MAP.
For reference, we also include the results from three computationally intensive cross-encoder models.
While cross-encoders (as expected) perform best due to their significant compute advantage, \modelname achieves comparable scores as a much more efficient bi-encoder model.
Our model's strong performance versus the RepLLaMA baseline illustrates that our instruction data is highly effective for dense retrievers, leading to significant gains in instruction following and prompt robustness.\footnote{It is likely that a cross-encoder trained on our generated data would outperform FollowIR-7B, however, we leave that for future work and focus only on dense retrievers in this work.}

\subsection{Standard Retrieval}
\label{sec:sec_standard_retrieval}
We benchmark \modelname on two standard retrieval tasks without instructions: both in-domain (MS MARCO) and out-of-domain (BEIR). In Table~\ref{tab:msmarco} we see that \modelname performs comparably to RepLLaMA on in-domain tasks despite additionally having  stronger instruction following performance.

\subsection{Retrieval with Prompts}
\label{sec:sec_with_few_shot_prompting}
A common approach to improving LMs on out-of-domain data is to include a textual prompt at test-time, even if the prompt is somewhat generic, e.g., ``think step by step" or ``I'll give you a tip."\cite{kojima2022large,wei2022chain,weller2023according} We apply this approach to IR by exploring whether particular prompts reliably induce improved retrieval performance in \modelname.

\begin{table}[t!]
\centering
\begin{tabular}{l|ccc}
\toprule
Model & DL19 & DL20 & Dev \\
 & \small nDCG@10 & \small nDCG@10 & \small MRR \\

\midrule
RepLLaMA & \textbf{74.5} & 71.8 & \textbf{42.5} \\
\modelname & 73.2 & \textbf{72.3} & 42.0 \\
\bottomrule
\end{tabular}
\caption{MS MARCO (in-domain) performance. We see that the models are comparable on all splits, despite \modelname being instruction-trained.\vspace{-1em}}
\label{tab:msmarco}
\end{table}

For out of domain performance on BEIR (Table~\ref{tab:beir}) without instructions (i.e. the \textit{None} column) we also see comparable scores: \modelname performs similarly to RepLLaMA (\modelname averaging 55.0 vs. 54.9 from RepLLaMA).

\begin{table*}[t!]
\setlength\tabcolsep{4 pt} %
\centering
\begin{tabular}{ll|ccc|ccc|ccc}
\toprule
& \multirow{2}{*}{Dataset} & \multicolumn{3}{c|}{BM25} & \multicolumn{3}{c|}{RepLLaMA} & \multicolumn{3}{c}{\modelname} \\
\cmidrule(l){3-5} \cmidrule(l){6-8} \cmidrule(l){9-11}
 & & \multirow{2}{*}{\shortstack{None}} & \multirow{2}{*}{\shortstack{Selected\\Prompt}} & \multirow{2}{*}{\shortstack{Best\\Prompt}} & \multirow{2}{*}{\shortstack{None}} & \multirow{2}{*}{\shortstack{Selected\\Prompt}} & \multirow{2}{*}{\shortstack{Best\\Prompt}} & \multirow{2}{*}{\shortstack{None}} & \multirow{2}{*}{\shortstack{Selected\\Prompt}} & \multirow{2}{*}{\shortstack{Best\\Prompt}} \\
 \\
\midrule
\parbox[t]{3mm}{\multirow{7}{*}{\rotatebox[origin=c]{90}{
\parbox[c]{3cm}{\centering \scriptsize Dataset has dev/train set}}}}
& DBPedia & 29.9 & 23.3 & 23.3 & 44.8 & 16.8 & 43.5 & 45.0 & \textbf{45.2} & \textbf{45.2} \\
& FEVER & 48.1 & 7.4 & 45.3 & 82.9 & \textbf{85.3} & \textbf{85.3} & 82.8 & 82.8 & 82.8 \\
& FiQA & 25.1 & 16.9 & 21.9 & 45.0 & 41.8 & 42.7 & 45.9 & \textbf{46.6} & \textbf{46.6} \\
& HotpotQA & 56.9 & 54.6 & 54.6 & 68.8 & 66.4 & 67.9 & 69.2 & \textbf{69.5} & \textbf{69.5} \\
& NFCorpus & 32.1 & 18.0 & 23.5 & 36.0 & 34.2 & 35.0 & 36.5 & \textbf{36.9} & \textbf{36.9} \\
& Quora & 80.4 & 75.3 & 75.3 & 86.0 & 83.1 & 85.5 & 86.5 & \textbf{88.0} & \textbf{88.0} \\
& SciFact & 68.7 & 65.7 & 65.7 & 75.3 & 75.0 & 75.7 & 75.0 & 75.9 & \textbf{76.3} \\
\midrule
\parbox[t]{5mm}{\multirow{6}{*}{\rotatebox[origin=c]{90}{
\parbox[c]{2.75cm}{\centering \scriptsize No dev/train set}}}}  & Arguana & 36.6 & - & 36.4 & 48.6 & - & 49.0 & 51.8 & - & \textbf{56.7} \\
& Climate-FEVER & 13.6 & - & 13.9 & 29.3 & - & 30.8 & 27.6 & - & \textbf{32.1} \\
& NQ\textsuperscript{\ref{foot:NQ}} & 28.5 & - & 25.4 & \textbf{63.0} & - & 62.2 & 61.9 & - & 62.6 \\
& SCIDOCS & 15.8 & - & 14.9 & 16.1 & - & 16.7 & 17.3 & - & \textbf{19.7} \\
& TREC-COVID & 62.3 & - & 35.9 & 83.9 & - & 82.7 & 83.9 & - & \textbf{84.6} \\
& Touche-2020 & 33.1 & - & 30.8 & 34.1 & - & \textbf{35.9} & 31.4 & - & 32.0 \\
\midrule
& Average & 40.9 & - & 35.9 & 54.9 & - & 54.8 & 55.0 & - & \textbf{56.4} \\
\bottomrule
\end{tabular}
\caption{Out of domain performance on BEIR (nDCG@10). \modelname performs similarly to RepLLaMA without any instructions (\textit{None} column). However, when given a prompt we see that performance improves for \modelname by +1.4 points whereas RepLLaMA and BM25 perform worse. \textbf{We thus see that \modelname is promptable due to its instruction-training. We also see that it is possible to select the best prompt from 10 dev examples consistently} for \modelname, as the difference between the \textit{Selected Prompt} is almost always the same as the \textit{Best Prompt} of the ten prompts evaluated. Note that not all BEIR datasets have training/dev sets and thus Selected Prompt is left blank for them. Best value in the row is bolded.\vspace{-1em}}
\label{tab:beir}
\end{table*}
\stepcounter{footnote}
\footnotetext{\label{foot:NQ}Although standard NQ has a training set, the BEIR NQ version selects a different set of queries and documents and doesn't provide a comparable dev/train set, c.f. \url{https://github.com/beir-cellar/beir/issues/179}.}

We use the following settings for testing prompts, following the standards in the LM community; typically one would evaluate prompts for an LM by first using a small validation set. We sample 10 queries from each of the validation (or train if there is no validation set) to use as the prompt tuning set. We also create 10 generic prompts\footnote{We include the generated prompts in Appendix~\ref{app:retrieval_prompts}} that could work across retrieval datasets.

However, not all of the BEIR datasets have train/dev data to sample validation examples from for selecting a prompt. We thus show results in two ways (Table~\ref{tab:beir}): (1) when there is a dev set: we select the best dev prompt as the test prompt (\textit{Selected Prompt} column) and leave the score blank for datasets without a dev/train set; and (2) taking the best prompt of the ten (\textit{Best Prompt} column).

We see in Table~\ref{tab:beir} that, for \modelname, using the best prompt brings significant gains to BEIR\footnote{Individual prompts and their scores on each BEIR dataset are found in Table~\ref{tab:prompt_ind_scores}.} average performance (+1.4 nDCG@10; gains versus no prompt for 12/13 datasets and tied on the last). However, in contrast, prompts fail to bring any gains to the RepLLaMA or BM25 models with -0.1 and -5.0 nDCG deltas respectively. Thus we can see that prompting is effective for \modelname but not for retrieval models using standard training.

\begin{table}[t!]
\small
\centering
\begin{tabular}{l|r|r|r}
\toprule
Dataset & BM25 & RepLLaMA & \modelname \\
\midrule
Arguana & 4.3 & \textbf{0.7} & 1.7 \\
C-FEVER & 4.8 & 2.4 & \textbf{2.3} \\
DBPedia & 9.0 & 9.3 & \textbf{2.2} \\
FEVER & 18.4 & 8.7 & \textbf{2.2} \\
FiQA & 6.1 & 4.6 & \textbf{2.8} \\
HotpotQA & 18.2 & \textbf{4.9} & 5.3 \\
NFCorpus & 7.9 & 3.8 & \textbf{3.7} \\
NQ & 7.8 & 9.8 & \textbf{2.1} \\
Quora & 23.9 & 7.1 & \textbf{0.9} \\
SCIDOCS & 4.5 & 2.0 & \textbf{1.0} \\
SciFact & 8.5 & \textbf{1.1} & 1.5 \\
TREC-COVID & 15.3 & 5.9 & \textbf{5.2} \\
Touche-2020 & 9.5 & \textbf{4.9} & 5.1 \\
\midrule
Average & 10.6 & 5.0 & \textbf{2.8} \\
\bottomrule
\end{tabular}
\caption{Standard deviations of NDCG@10 scores per dataset across all prompts. Best value in the row is bolded. Note that absolute scores for BM25 and RepLLaMA are lower, but some RepLLaMA scores have lower variance due to clustering at these lower scores.\vspace{-1em}}
\label{tab:std_devs}
\end{table}

\begin{table*}[t]
\centering
\resizebox{\textwidth}{!}{%
\begin{tabular}{l|cc|cc|cc|cc|cc}
\toprule
 \multirow{3}{*}{Model} & \multicolumn{8}{c|}{FollowIR} & \multicolumn{2}{c}{InstructIR}  \\
 & \multicolumn{2}{c|}{Robust04} & \multicolumn{2}{c|}{News21} & \multicolumn{2}{c|}{Core17} & \multicolumn{2}{c|}{Average} & \multicolumn{2}{c}{MS MARCO}  \\
 & MAP & p-MRR & nDCG & p-MRR & MAP & p-MRR & Score & p-MRR & nDCG & Robust. \\
 \midrule
  RepLLaMA & 24.0 & -8.9 & 24.5 & -1.8 & 20.6 & +1.3 & 23.0 & -3.1 & 85.7 & 50.2 \\
  \midrule
Repeated Query & 24.6 & -9.1 & 25.3 & -2.6 & 21.1 & +2.4 & 23.6 & -3.1 & 85.4 & 49.2 \\
 Generic Instruct & 25.5 & -7.2 & 26.2 & -1.7 & 21.6 & -0.0 & 24.4 & -3.0 & 63.1 & 32.4 \\
 Swap Instruct & 25.2 & -1.9 & 27.3 & -0.2 & 21.1 & -0.6 & 24.6 & -0.9 & 48.6 & 27.0 \\
w/Instructions & 26.9 & +3.8 & \textbf{29.1} & +5.3 & 20.7 & +8.0 & 25.6 & +5.7 & 91.9 & \textbf{63.3} \\
w/Instruction Negatives & \textbf{29.0} & +9.7 & 27.8 & +5.2 & \textbf{21.9} & +11.4 & \textbf{26.2} & +8.8 & 91.5 & 62.0 \\
\midrule
\modelname (Joint) & 28.3 & \textbf{+11.7} & 28.5 & \textbf{+6.4} & 21.6 & \textbf{+15.4} & 26.1 & \textbf{+11.2} & \textbf{92.1} & 63.1 \\

\bottomrule
\end{tabular}
}
\caption{Ablations for instruction following on the FollowIR and InstructIR datasets. \textbf{The instruction provides gains beyond the simple length of the instruction or its distribution}. Instruction-negatives and joint data bring even more gains. Best value is in the column is bolded.\vspace{-1em}}
\label{tab:instruct_ablate}
\end{table*}

But are these best prompt numbers close to what would be achieved in practice with a small dev set? The \textit{Selected Prompt} column shows score from a practical setting.
If we compare \modelname's score for Selected Prompt vs Best Prompt, we see that there is very little difference between them. Applying few-shot selection with the dev set selects the best prompt in 6/7 cases, and in the 7th case (SciFact) it chooses a prompt that is still almost one full point better than the no prompt setting.
In contrast, and as expected, BM25 is not ``promptable" in any setting with performance dropping across the board; RepLLaMA's performance drops (sometimes dramatically) in six of seven cases. %

We also examined the sensitivity of all models to the prompts (Table~\ref{tab:std_devs}).
We see that \modelname's variance to prompts is significantly less than that of RepLLaMA (by 44\%) and BM25 (by 77\%) which has wide swings due to the effect of the keyword matching.
This suggests that \modelname is more robust to the input as well.

In summary, the standard practice of instruction-training with LMs can apply to instruction-trained dense retrievers as well, but only if they, like \modelname, are trained to be sensitive to such prompts. Furthermore, strong gains are indeed reliably possible to achieve via selecting a natural language prompt using a small held-out eval set.

\section{Analysis}
\label{sec:ablations}
We ablate several null hypotheses in an effort to better understand which parts of the \modelname training recipe contribute most to performance gains. We train all of the ablated models on a dataset consisting of half MS MARCO data and half instruction-data for compute and speed reasons (also matching the number of training instances in RepLLaMA). Results for all ablations are in Table~\ref{tab:instruct_ablate}, with the baseline being RepLLaMA, and each row of the table representing either a null hypothesis or a design decision leading to the final \modelname models.

\begin{table*}[t]
\centering
\resizebox{\textwidth}{!}{%
\begin{tabular}{ll|ccc|cc|cc|cc}
\toprule
& \multirow{2}{*}{Base Model} & \multicolumn{3}{c|}{MS MARCO} & \multicolumn{2}{c|}{BEIR} & \multicolumn{2}{c|}{FollowIR} & \multicolumn{2}{c}{InstructIR} \\
 \cmidrule(l){3-5} \cmidrule(l){6-7} \cmidrule(l){8-9} \cmidrule(l){10-11}
 & & DL19 & DL20 & Dev & nDCG & w/Prompt & Score & p-MRR & nDCG & Robust@10 \\
 \midrule
&  RepLLaMA & 73.2 & 72.3 & 42.0 & 54.9 & 54.8 & 23.0 & -3.1 & 85.7 & 50.2 \\
\midrule
\parbox[t]{2mm}{\multirow{4}{*}{\rotatebox[origin=c]{90}{
\parbox[c]{1.9cm}{\centering \small Promptriever}}}} & LLaMA2 & 74.5 & 71.8 & 42.5 & 55.0 & 56.4 & 26.1 & +11.2 & 92.1 & 63.1 \\
& Mistral v1 & 72.9 & 73.4 & 42.3 & 54.4 & 55.7 & 25.7 & +11.8 & 90.3 & 58.8  \\
& Llama 3.1  & 73.5 & 72.9 & 43.2 & 55.1 & 56.5 & 25.0 & +11.3 &  85.5 & 41.6 \\
& Llama 3.1 Instruct & 72.4 & 73.6 & 42.7 & 55.5 & 57.2 & 26.0 & +9.8 & 89.9 & 57.8   \\
\bottomrule
\end{tabular}
}
\caption{Comparison of different backbone models on the same \modelname recipe across MS MARCO datasets (DL19, DL20, and Dev), BEIR, InstructIR, and FollowIR. \textbf{We see that our augmented instruction dataset provides gains for many different base models, indicating the generality of our approach}. Hyperparameter tuning would likely improve these results for other backbone models.}
\label{tab:model-benchmark-comparison}
\end{table*}

\paragraph{Q1: Is it simply the length of the query/instruction that enables \modelname's performance gains?} \textbf{\emph{Answer: No.}} We train models with the query repeated to the length of the instruction (\textit{Repeat Query}) and with \textit{Generic Instructions.}\footnote{We add a generic retrieval instruction from one of 50 different generic retrieval task descriptions generated by \texttt{GPT-4o} and \texttt{Claude-3.5-Sonnet}. See a full list in Appendix~\ref{app:generic}.} Increasing the length of RepLLaMA queries results in a slight gain on standard retrieval performance, but little gain in p-MRR (retrieval sensitivity). This includes the \textit{Repeat Query} (+0.6 nDCG/MAP) and \textit{Generic Instruction} (+1.4 nDCG/MAP) baselines. 

\paragraph{Q2: Is it the lexical distribution of the instructions (in isolation) that enables these gains?} \textbf{\emph{Answer: Partially.}} For this we train with the real generated instructions but randomly swap the instruction each query is paired with (\textit{Swap Instructions} row). Compared to the length ablations, we see larger gains in the score (+1.6) and p-MRR (+2.1) as the model has learned the distribution, although not how to use them effectively. 

\paragraph{Q3: How much does training with instructions help?}
\textbf{\emph{Answer: Significantly.}} We ablate this by showing the results of training with just the instructions and no instruction-negatives (\textit{w/Instructions}). We see a strong gain in p-MRR (+6.6) and a further gain in standard retrieval (+1) over Swap.

\paragraph{Q4: How much does training with instruction-negatives help?}
\textbf{\emph{Answer: Significantly.}}
Adding the instruction negatives on top of \textit{w/Instructions} gives another large gain in p-MRR (+3.1 over \textit{w/Instructions}) and a small boost in standard retrieval scores (+0.6 nDCG/MAP). This aligns with expectations: instruction negatives provide extra data for instruction sensitivity but not necessarily for standard retrieval metrics.

\paragraph{Q5: Does training additionally on MS MARCO help beyond the \modelname training set we curate?}
\textbf{\emph{Answer: Yes.}}
\emph{\modelname (Joint)} our final model, combines all the MS MARCO and Instruction data which leads to another large jump in p-MRR (+2.4) as it is able to see more data (and instructions) in training, i.e. 2x as much. 

\paragraph{In summary,} each step in our final recipe (+w/ Instructions, +w/ Instruction Negatives, +MS MARCO Jointly) provides value independently, and that value is not due to simple factors like increasing the length of the query and/or surface lexical features of the instructions.

\subsection{Does this process work for other models?}
\label{sec:backbones}
The original RepLLaMA used Llama 2 as a backbone, and, to this point in our paper, \modelname has also used Llama 2 as a backbone for fair comparison. 
We also adopt the same training hyperparameters as RepLLaMA.
Nonetheless, we ablate different LM backbones to see if performance holds without any adjustments to the hyperparameters or training recipe.
While further tuning the learning rate and other parameters would likely improve performance, %
we see in Table~\ref{tab:model-benchmark-comparison} that other backbones provide comparable performance, indicating the generality of our method.

\section{Related Work}
\subsection{Instructions in Retrieval}

The use of instructions %
is a relatively new development for IR models, as
dense retriever training generally focuses on learning similarity functions similar to phrase-level matching \cite{craswell2020overview,izacard2021unsupervised,wang2022text}. %
Some of the earliest work on the topic is TART \cite{asai2022tart} and Instructor \cite{instructor_models} which used simple task prefixes during training. More recently, %
E5-Mistral \cite{wang2023improving}, GritLM \cite{muennighoff2024generative}, and NV-Retriever \cite{moreira2024nvretrieverimprovingtextembedding} %
scaled up the dataset and model size. These newer models typically re-use the same instruction set proposed by the E5-Mistral model.\footnote{You can find a list of all the ``instructions'' \href{https://github.com/microsoft/unilm/blob/master/e5/utils.py\#L163}{at this url}.}
Our work differs from this by applying and evaluating adaptability \emph{per-query} rather than using a dataset wide prefix.

Our robustness evaluation also goes beyond prior works: while, e.g., \citet{instructor_models} tests %
instruction phrasing by changing one word, we consider a broader range of length/style modifications.

Several benchmark efforts focus on explicitly testing the instruction following ability of retrievers: FollowIR \cite{weller2024followir} and InstructIR \cite{oh2024instructir}. Both found that existing bi-encoder retrieval models fail to use instructions as an LM would. Our work presents the first bi-encoder that achieves significantly above-random performance on these benchmarks. %

\subsection{Prompting LMs}
It is now the defacto-standard for LMs to take and reason over input instructions given via \textit{prompting}. This was discovered and popularized by models such as InstructGPT~\citep{instructGPT}, FLAN~\citep{wei2022finetuned}, and T0~\citep{sanh2022multitask}. Importantly, these works found that diversity of training data was crucial to generalization.
Instructions are also often included in LM's training data to encourage this behavior, both in pre-training data \cite{soldaini2024dolma,together2023redpajama} and followed by stages of post-training (including fine-tuning/RL \citet{groeneveld2024olmo}).

Although IR models often use LMs as their base architecture before IR training, little work has explored using standard LM capabilities like promptability in IR. The closest works include GritLM \cite{muennighoff2024generative} who attempted to do in-context learning (ICL) with their trained retriever but found worse results than zero-shot and potentially the recent BGE-ICL,\footnote{\url{https://huggingface.co/BAAI/bge-en-icl}} although as of publication, there is no associated paper describing their procedure or training details. 

\subsection{Synthetic Instruction Generation}
Generating synthetic data for training is a popular technique given the strong performance of LMs. This happens in both NLP \cite{benallal2024cosmopedia,adler2024nemotron} as well as in IR, for generating queries or documents for training \cite{bonifacio2022inpars,dai2022promptagator,jeronymo2023inpars,Weller2024NevIRNI}.

We build upon another line of work that uses LMs to create instructions that can be used for retrieval training %
\cite{wang2022self,li2023self,chung2024scaling} which we use to generate instruction-negative passages.

\section{Conclusion}
We presented the first zero-shot \textit{promptable} retriever, \modelname, trained from a new instruction-based retrieval dataset based on MS MARCO. Experiments show that \modelname not only performs well on the standard retrieval task, but also follows instructions more effectively than prior work, adapting its notion of relevance per-query. %
Overall, this shows that techniques discovered in the LM community, such as prompting, can be extended to dense retrievers as well. We hope this will inspire joint research between the two communities and further enable controllable retrievers that can adapt on the fly.

\section{Limitations}
Although \modelname introduces per-instance prompting to retrieval models, there are many aspects of prompting with LMs that we did not explore in this work. For example, future work could look at in-context learning: can retrieval models be prompted with a few examples explicitly versus just imperative requests? %

We also note that, similar to language models, it is often unclear why some IR prompts perform better than others. Language models have become more robust to different prompts over time, and we hope that future work will continue to improve this ability for retrieval models.  

Finally, as with any LM-generated data, it is possible that there are errors, pernicious social biases, and/or incorrect pieces of information in the generated passages and instructions. Although we applied some (probabilistic) correctness filters and conducted quantitative+qualitative explorations, it is still possible that unintended characteristics slipped through. While experiments demonstrate that training on our corpus improves performance, further audits of our corpus (and retrieval training sets more broadly) would be appropriate.

\bibliography{custom}

\appendix

\section{Retrieval Prompts}
\label{app:retrieval_prompts}
In Table~\ref{tab:generic_prompts} we show the retrieval prompts and their scores per dataset in Table~\ref{tab:prompt_ind_scores}.

\section{Hyperparameter Details}
\label{app:hyperparameters}
We use the following hyperparameters as given by the authors of the RepLLaMA on their \href{https://github.com/texttron/tevatron/issues/129}{Github page} using Tevatron \cite{gao2022tevatron}. This is using \texttt{meta-llama/Llama-2-7b-hf} with lora r 32, lora modules q\_proj, k\_proj, v\_proj, o\_proj, down\_proj, up\_proj, gate\_proj, enabled bfloat16, using eos pooling, using normalization, a temperature of 0.01, learning rate of 1e-4, one epoch, passage length 256, 100 warm up steps, a train group size of 16, and an effective batch size of 128 (4 GPUs, 8 per device with a 4 accumulation steps). We differ from the original paper by using query max length 304 to account for the longer instructions (previously set to 32 in RepLLaMA). Training takes approximately 2 days on 8x40GB A100s for the ablation runs and 4 days for the full run.  Inference takes up to four hours per dataset on the 8x cluster using 512 length parameters for query and passages.

\section{Generation Prompts}
\label{app:prompts}
We include the prompts used to generate the data. For the system prompts, we used Prompt~\ref{fig:system_prompt}, for the instruction generation we used Prompt~\ref{fig:instruction_prompt}, and for the instruction negatives we used Prompt~\ref{fig:hard_positive_prompt}.

\begin{figure*}
\begin{tcolorbox}[
    colback=white,
    colframe=black,
    title=Prompt for Instruction Generation
    ]
\#\# \textbf{Input Data}

I have the following query and REL\_DOCS\_NUM\_FILL\_ME documents which have been marked as relevant and NON\_REL\_DOCS\_NUM\_FILL\_ME which are non-relevant.

\medskip

Query: QUERY\_FILL\_ME

\medskip

POS\_DOC\_FILL\_ME

\medskip

NEG\_DOC\_FILL\_ME

\medskip

\#\# \textbf{Your task}

I need you to come up with an instruction that can be appended onto the end of this query that will \textbf{make only one relevant document} and \textbf{make all other documents (including previously relevant docs) non-relevant}. You can choose which document will stay relevant to the new instruction, by writing an instruction that applies to only one of the relevant documents (you choose). This additional instruction should provide a test for strong frontier language models to determine if they can follow instructions. Triple check your work to be certain that the chosen document is still relevant and that the others are non-relevant -- if you mess up you will be fired. Do not give away the answer in the instruction!

\medskip

For this example, please generate the instruction to be LENGTH\_FORMAT\_FILL\_ME. \textbf{In the instructions, provide detailed specifics for what makes a document relevant.} Remember that this criteria should make the one document relevant and all others irrelevant. Also be sure that the \textbf{instruction is generic and does not contain the answer to the query}.

\medskip

Output the response in JSON form only with no other text, with the keys, ``instruction'' (str), ``relevant\_docs'' (one document id that is the first doc, e.g. ``[2]'') and ``non-relevant\_docs'' (all other document ids, e.g. ``[1,3,...]'').

\medskip

\#\# \textbf{Your output (JSON only):}
\end{tcolorbox}
\caption{Prompt for Instruction Generation}
\label{fig:instruction_prompt}
\end{figure*}

\begin{figure*}
\begin{tcolorbox}[colback=white,colframe=black,title=Prompt for instruction negatives]
Generate three ~100 word passages (and explanations) that directly answer the query but do not provide a valid document according to the specific query. Then generate one passage that matches both. Make it obvious to a reader which ones are which.

\medskip
\medskip

Query: QUERY\_FILL\_ME

\medskip

Specific Query: INSTRUCTION\_FILL\_ME

\medskip

Remember your goal is to \textbf{generate a relevant document (MS MARCO style, with passage and title) for the query} but a \textbf{non-relevant document for the specific query}. You should generate only factual information.

\medskip

To be crystal clear, your generated documents should have related information about "QUERY\_FILL\_ME". However, these generated documents should not be relevant to the specific query. As examples, they may omit crucial information that is needed for the specific query, if the query is ambiguous it may use an alternative meaning, or it may specifically mention elements that are said to be non-relevant.

\medskip

You should also generate an explanation with the category it is and a succinct reason. The tags are "different interpretation", "omission", "mention non-relevant flag" or "none" for the relevant to both. E.g. "omission - it does not mention [reason]". \textbf{Be sure the documents marked as non-relevant to both are actually not relevant to the specific query!!}

\medskip

Remember! \textbf{It should be trivially obvious to a reader why they are non-relevant!}

\medskip

Diverse Generated Documents in \textbf{JSON output} with "matches\_both", "explanation", "title", and "passage" keys. Reply with only valid JSON, no other text:
\end{tcolorbox}
\caption{Prompt for Instruction Negatives}
\label{fig:hard_positive_prompt}
\end{figure*}

\begin{figure}
\begin{tcolorbox}[
  colback=white,
  colframe=black,
  title={System Prompt for All Prompts}]
You are an expert at writing precise detailed instructions for language models and are paid millions of dollars to be a data engineer for OpenAI. Your sole duty is to write instructions that can be used for training data for the next superpowerful model, GPT-6. Answer succinctly and carefully follow all instructions given so that you can earn your large bonus and not be fired.
\end{tcolorbox}
\caption{System Prompt for All Prompts}
\label{fig:system_prompt}
\end{figure}

\section{Generic Instructions}
\label{app:generic}
We show the generic instructions given the to the models in Table~\ref{tab:generic_instructions}. These were generated by prompting GPT-4o and Claude-3.5-Sonnet for generic retrieval descriptions.

\section{Error Analysis Attempts}
In order to further understand why \modelname was more effective than RepLLaMA we conducted the following error analyses. For the datasets with the largest differences between the two (and for the difference between \modelname with prompt and without prompt, such as Climate-FEVER, SciFact, and Arguana) we calculated the per-query nDCG@10 scores. We then binned the queries into those that saw improved performance vs those that did not.  Finally, we fine-tuned a BERT-base model and a bag-of-words model on 80\% of those examples, leaving 20\% for a hold out test set. However, in every case we found that the accuracy was below the majority baseline (typically around 66\%).  The best AUC score we found for any dataset was 54\%, further indicating that there was not much signal in the data. We attempted several other combinations (including adding queries with tied scores in the negative bin, adding the prompt text to the queries) none of which changed these results. We hypothesize that the documents must be a critical component to understanding why \modelname works better or why the prompts are helpful (or alternatively, the query-document connection). 

We further tried simple statistics of the binned queries but found the two groups were indentical w.r.t. length, idf, and other basic text statistics.

\section{Comparison to state-of-the-art models on BEIR}
Table~\ref{tab:beir_comp} compares \modelname to some of the best models\footnote{Details for Voyage's model is \href{https://blog.voyageai.com/2024/05/05/voyage-large-2-instruct-instruction-tuned-and-rank-1-on-mteb/}{here} and TDTE \href{https://github.com/raghavlite/TDTE}{here}.} \cite{muennighoff2024generative,wang2023improving,lee2024gecko,behnamghader2024llm2vec,merrick2024arctic,li2023towards} on the BEIR leaderboard \cite{muennighoff2022mteb} (from MTEB). We note that our model does not train on any of the BEIR datasets except for MS MARCO, which puts it at a disadvantage (as most SOTA models use all training/dev sets). Despite this, we see solid performance, including middle-of-the-pack performance when compared on only datasets without train/dev sets (truly OOD performance) beating GritLM, LLM2Vec, and Google Gecko.

\begin{table*}[htbp]
\centering
\begin{tabular}{p{0.9\textwidth}}
\hline
\textbf{Generic Instructions} \\
\hline
\textbullet~Retrieve relevant passages. \\
\textbullet~Find answer-containing text. \\
\textbullet~Rank based on relevance. \\
\textbullet~Identify key information. \\
\textbullet~Extract pertinent information. \\
\textbullet~Rank documents based on query relevance. \\
\textbullet~Retrieve passages that answer the user's question. \\
\textbullet~Find relevant passages. \\
\textbullet~Rank matching documents. \\
\textbullet~Retrieve answer-containing text. \\
\textbullet~Identify key information sources. \\
\textbullet~Find and rank passages that best address the user's query. \\
\textbullet~Locate text segments that are relevant to the query and rank them. \\
\textbullet~Identify and retrieve passages that answer the user's question. \\
\textbullet~Extract passages from the corpus that are most relevant to the given query. \\
\textbullet~Rank passages based on their ability to address the core aspects of the query. \\
\textbullet~Given a web search query, retrieve relevant passages that answer the query. \\
\textbullet~Find relevant passages for the given query. \\
\textbullet~Select the most relevant passages that directly answer this query. \\
\textbullet~Rank documents based on their relevance and informativeness to the given question. \\
\textbullet~Retrieve passages containing factual information that addresses this specific inquiry. \\
\textbullet~Identify and rank sources that provide comprehensive answers to the posed question. \\
\textbullet~Analyze the query to identify the key information needs. Retrieve and rank passages that provide comprehensive answers to those needs. \\
\textbullet~Locate relevant passages that directly respond to the user's question. Ensure the passages are ranked based on their relevance and accuracy. \\
\textbullet~Search for text that addresses the user's query. Rank the passages based on how well they meet the information needs and provide clear answers. \\
\textbullet~Examine the query for specific details and retrieve passages that address those details. Rank the results by their relevance and comprehensiveness. \\
\textbullet~Extract pertinent information from the corpus to address the given query. \\
\textbullet~Locate and prioritize text segments that provide accurate answers to the user's question. \\
\textbullet~Evaluate document relevance based on query similarity and information content. \\
\textbullet~Identify passages containing key facts related to the input query. \\
\textbullet~Parse the query, then retrieve and rank relevant textual information. \\
\hline
\end{tabular}
\caption{Generated Generic Instructions for IR, as generated by GPT-4o and Claude-3.5-Sonnet in July 2024. Prompts asked the models to generate them of varying length.}
\label{tab:generic_instructions}
\end{table*}

\begin{table*}[htbp]
\centering
\begin{tabular}{lp{0.7\textwidth}}
\toprule
\textbf{Type} & \textbf{Example} \\
\midrule
Negation & A relevant document is one that provides information about a specific city or town, including its location, population, and history. It should not be about a trail, a business, or a resort. The document should also contain specific details about the city or town, such as its county or state. \textbf{Documents that only mention the name of the city or town in passing are not relevant.} \\
\midrule
Persona & \textbf{I'm a history teacher preparing a lesson on the origins of the Pledge of Allegiance} and I need documents that provide a clear and specific answer to when it was written, including the name of the author and their occupation. A relevant document should provide a direct quote or explicit statement about the creation of the Pledge. \\
\midrule
Background & \textbf{In the field of chemistry, substances can be classified into different categories based on their composition and properties. A thorough understanding of these categories is essential to accurately identify and describe various substances.} When evaluating a document's relevance to the question of whether gasoline is a substance or mixture, consider the following criteria: a relevant document must explicitly address the composition of gasoline, discussing its homogeneity or heterogeneity, and provide specific details about its properties or behavior under different conditions. The document should also demonstrate a clear understanding of the distinction between substances and mixtures, and apply this understanding to the case of gasoline. Furthermore, a relevant document should not simply provide a general definition of a substance or mixture, but rather provide specific information about gasoline that helps to answer the question \\
\bottomrule
\end{tabular}
\caption{Examples of instruction features in our new dataset, including negation, a POV, and background information.}
\label{tab:instruction_examples}
\end{table*}

\begin{table*}[htbp]
\centering
\begin{tabular}{lp{0.7\textwidth}}
\toprule
\textbf{Type} & \textbf{Example} \\
\midrule
Short (1-2 sentences) & Documents that describe the authorship of a specific hymn or song, mentioning the writer's name and the song's title, are relevant. Documents that discuss general information about bands, poems, or biblical events are not relevant. \\
\midrule
Medium (3-6 sentences) & Proton pump inhibitors are a class of medications that have been widely used for several decades. They are available both over-the-counter and by prescription. A relevant document should provide a clear explanation of how proton pump inhibitors work to reduce stomach acid, and specifically mention their effect on the body's production of stomach acid. The document should also discuss the medical conditions that proton pump inhibitors are used to treat. A relevant document should not simply list the names of proton pump inhibitors or their uses without providing a detailed explanation of their mechanism of action. \\
\midrule
Long (one paragraph) & Nicotine is a highly addictive substance found in tobacco products, and its detection in the body is a crucial aspect of medical testing. The human body has various ways of eliminating nicotine, including through urine, blood, and hair follicles. When evaluating documents related to nicotine detection, it is essential to consider the specific context and criteria for relevance. A relevant document should provide a clear and concise answer to the question, specifying the duration of nicotine presence in the body, particularly in urine tests. The document should also discuss the factors that influence nicotine detection, such as the frequency and amount of smoking, as well as the role of passive smoking. Furthermore, a relevant document should provide a comprehensive overview of nicotine's effects on the body and its elimination process. Documents that merely list detection periods without providing a detailed explanation of the underlying factors or fail to address the specific context of urine tests should be considered non-relevant. \\
\midrule
Very Long (two paragraphs) & I'm planning a road trip from Red Lodge to Cooke City, Montana, and I'm looking for information on the route that will take me through the most scenic and thrilling parts of the Montana-Wyoming border. I've heard that there's a particular highway that's known for its steep switchbacks and breathtaking views, and I want to know more about it. A relevant document would need to provide specific details about the highway, such as its name, elevation gain, and any notable features or landmarks along the way. It's crucial that the document focuses on the highway itself, rather than general information about road trips or travel in Montana and Wyoming. 

I'm not interested in documents that talk about motorcycle helmets, natural arches, or teething symptoms - those are completely unrelated to my road trip plans. A relevant document should make me feel like I'm getting a firsthand account of the highway and its attractions.I've tried searching online, but I keep getting results that are either too vague or too focused on other aspects of travel. That's why I need a document that can provide me with the specific information I'm looking for. If a document can give me a clear sense of what to expect on this highway, including its length, elevation, and any notable features, then I'll know it's the right one. Anything less, and I'll have to keep searching. \\
\bottomrule
\end{tabular}
\caption{Examples of instructions by length format.}
\label{tab:length_examples}
\end{table*}

\begin{table*}[htbp]
\centering
\begin{tabular}{p{0.9\textwidth}}
\hline
\textbf{Prompts} \\
\hline
\textbullet~Be careful when assigning relevance as your job is on the line and I will give you a 1000 dollar tip. \\
\textbullet~Think carefully about these conditions when determining relevance. \\
\textbullet~A relevant document should also provide a clear and concise explanation, avoiding unnecessary complexity or ambiguity. When in doubt, prioritize documents that provide a clear, direct, and specific answer to the query. \\
\textbullet~A document that meets these criteria is considered relevant, while a document that does not meet these criteria is considered non-relevant. \\
\textbullet~A relevant document should focus solely on providing a clear and accurate answer to the query, without distracting or unnecessary information \\
\textbullet~A document is relevant if it helps to answer the query. Surface relevant documents only. \\
\textbullet~Relevant documents are those that are topically related, answer the given question, or otherwise provide insight on the input. Think step by step about whether a document is relevant for this question. \\
\textbullet~Find relevant documents to the query. Use strict critera when evaluating relevance: a relevant document here should provide direct information to either fully answer the query, or provide useful information towards answering it. Avoid only topically relevant documents. \\
\textbullet~When judging the relevance of a document, focus on the pragmatics of the query and consider irrelevant any documents for which the user would have used a different query. \\
\textbullet~Think carefully about relevance \\
\hline
\end{tabular}
\caption{Prompts used for BEIR experiments. Results for each dataset is shown in Table~\ref{tab:prompt_ind_scores}.}
\label{tab:generic_prompts}
\end{table*}

\begin{table*}[h]
\tiny
\setlength\tabcolsep{2 pt} 
\begin{tabular}{p{8.25cm}|ccccccccccccc}
\toprule
Prompt & ARG & CFV & DBP & FEV & FQA & HQA & NFC & NQ & QUO & SCD & SCF & COV & TOU \\
\midrule
\tiny{A relevant document should focus solely on providing a clear and accurate answer to the query, without distracting or unnecessary information} & 55.9 & 29.4 & 43.7 & 78.1 & 42.2 & 68.0 & 35.9 & 58.1 & \textbf{88.0} & 19.3 & 75.9 & 73.3 & 19.2 \\
\tiny{A relevant document should also provide a clear and concise explanation, avoiding unnecessary complexity or ambiguity. When in doubt, prioritize documents that provide a clear, direct, and specific answer to the query.} & \textbf{56.7} & \textbf{32.1} & 43.1 & 77.3 & 38.7 & 67.5 & 35.0 & 56.1 & 87.2 & \textbf{19.7} & 75.0 & 64.7 & 18.3 \\
\tiny{Think carefully about relevance} & 52.7 & 27.5 & 44.8 & \textbf{82.8} & 43.7 & \textbf{69.5} & 36.5 & 61.5 & 85.9 & 17.8 & 76.2 & 82.5 & \textbf{32.0} \\
\tiny{A document that meets these criteria is considered relevant, while a document that does not meet these criteria is considered non-relevant.} & 51.4 & 26.4 & \textbf{45.2} & 80.1 & \textbf{46.6} & 69.0 & \textbf{36.9} & \textbf{62.2} & 86.8 & 18.3 & 74.9 & \textbf{84.6} & 30.4 \\
\tiny{Think carefully about these conditions when determining relevance.} & 53.2 & 26.7 & 44.9 & 81.9 & 43.1 & 69.3 & 35.9 & 61.1 & 84.9 & 18.0 & \textbf{76.3} & 81.5 & 30.2 \\
\tiny{When judging the relevance of a document, focus on the pragmatics of the query and consider irrelevant any documents for which the user would have used a different query.} & 53.3 & 24.0 & 43.1 & 78.5 & 43.4 & 68.3 & 34.3 & 59.4 & 86.6 & 18.0 & 75.1 & 79.9 & 27.8 \\
\tiny{Find relevant documents to the query. Use strict critera when evaluating relevance: a relevant document here should provide direct information to either fully answer the query, or provide useful information towards answering it. Avoid only topically relevant documents.} & 51.5 & 26.2 & 44.4 & 79.3 & 45.3 & 67.3 & 36.6 & 60.0 & 87.2 & 18.5 & 75.2 & 82.4 & 30.4 \\
\tiny{Relevant documents are those that are topically related, answer the given question, or otherwise provide insight on the input. Think step by step about whether a document is relevant for this question.} & 54.6 & 27.0 & 43.8 & 80.9 & 41.2 & 68.9 & 35.4 & 57.2 & 86.6 & 17.6 & 76.1 & 77.9 & 24.1 \\
\tiny{A document is relevant if it helps to answer the query. Surface relevant documents only.} & 52.9 & 27.3 & 43.4 & 79.4 & 45.0 & 67.3 & 35.7 & 58.6 & 86.8 & 17.9 & 75.5 & 79.8 & 28.5 \\
\tiny{Be careful when assigning relevance as your job is on the line and I will give you a 1000 dollar tip.} & 52.4 & 24.3 & 43.2 & 81.0 & 41.0 & 68.6 & 35.5 & 60.5 & 84.4 & 18.3 & 75.4 & 83.6 & 25.4 \\
\bottomrule
\end{tabular}
\caption{BEIR dataset scores for different prompts (shown larger in Table~\ref{tab:generic_prompts}) for \modelname}
\label{tab:prompt_ind_scores}
\end{table*}

\begin{table*}[htbp]
\centering
\begin{tabular}{@{}r|rrrrrrrrr@{}}
\toprule
Dataset & \rotatebox{90}{GritLM-7B} & \rotatebox{90}{e5-mistral-7b-instruct} & \rotatebox{90}{voyage-lite-02-instruct} & \rotatebox{90}{gte-Qwen1.5-7B-instruct} & \rotatebox{90}{google-gecko} & \rotatebox{90}{LLM2Vec-Mistral-supervised} & \rotatebox{90}{snowflake-arctic-embed-l} & \rotatebox{90}{Promptriever-llama2-7b} & \rotatebox{90}{TDTE} \\
\midrule
\multicolumn{10}{l}{\textbf{Has train/dev set}} \\
DBPedia & 46.6 & 48.9 & 39.8 & 48.0 & 47.1 & 49.6 & 46.0 & 45.2 & 53.2 \\
FEVER & 82.7 & 87.8 & 91.4 & 93.4 & 87.0 & 89.4 & 88.2 & 82.8 & 77.7 \\
FiQA2018 & 60.0 & 56.6 & 52.5 & 55.3 & 59.2 & 53.1 & 44.7 & 46.6 & 40.7 \\
HotpotQA & 79.4 & 75.7 & 75.5 & 72.3 & 71.3 & 74.1 & 75.2 & 69.5 & 41.3 \\
NFCorpus & 40.9 & 38.6 & 43.7 & 38.3 & 40.3 & 39.3 & 37.7 & 36.9 & 88.9 \\
NQ & 70.3 & 63.5 & 64.3 & 61.8 & 61.3 & 61.7 & 63.1 & 62.6 & 23.0 \\
Quora Retrieval & 89.5 & 89.6 & 87.6 & 89.6 & 88.2 & 87.8 & 87.4 & 88.8 & 79.6 \\
SciFact & 79.2 & 76.4 & 79.9 & 75.3 & 75.4 & 78.9 & 73.8 & 76.3 & 80.8 \\
\midrule
\multicolumn{10}{l}{\textbf{Doesn't have train/dev set}} \\
ArguAna & 63.2 & 61.9 & 70.3 & 62.7 & 62.2 & 57.5 & 59.1 & 56.7 & 49.5 \\
Climate FEVER & 30.9 & 38.4 & 32.0 & 44.0 & 33.2 & 35.2 & 39.3 & 32.1 & 49.0 \\
SCIDOCS & 24.4 & 16.3 & 20.2 & 27.7 & 20.3 & 22.5 & 21.4 & 19.7 & 25.2 \\
Touche2020 & 27.9 & 26.4 & 26.8 & 20.3 & 25.9 & 22.2 & 34.5 & 32.0 & 22.0 \\
TRECCOVID & 74.8 & 87.3 & 81.0 & 72.7 & 82.6 & 77.7 & 80.7 & 84.6 & 58.8 \\
\midrule
Average & 59.2 & 59.0 & 58.8 & 58.6 & 58.0 & 57.6 & 57.8 & 56.4 & 53.1 \\
Average OOD & 44.3 & 46.0 & 46.1 & 45.5 & 44.8 & 43.0 & 47.0 & 45.0 & 47.0 \\
\bottomrule
\end{tabular}
\caption{BEIR comparison for models in the MTEB leaderboard. \modelname, unlike most others, has not been trained on the training/dev sets of the BEIR datasets (other than MS MARCO). Despite that, it performs comparably to many models on the true out-of-distribution (OOD) datasets that don't have train/dev sets.}
\label{tab:beir_comp}
\end{table*}

\end{document}